\documentclass[runningheads]{llncs}
\usepackage{graphicx}
\usepackage[shortcuts, acronym]{glossaries}
\usepackage{xcolor}
\usepackage{placeins}
\usepackage{booktabs}
\usepackage{array}
\usepackage[english]{babel}
\usepackage{hyperref} 
\usepackage{multirow}
\usepackage{amssymb} 
\usepackage{circledsteps}
\usepackage[font=small,labelfont=bf]{caption} 
\usepackage{subcaption} 

\pgfkeys{/csteps/inner color=white}
\pgfkeys{/csteps/outer color=black}
\pgfkeys{/csteps/fill color=black}
\definecolor{mygreen}{RGB}{0, 120, 0}

\newcolumntype{M}[1]{>{\centering\arraybackslash}m{#1}}
\newcolumntype{Z}[1]{>{\raggedright}m{#1}}

\newacronym{gpu}{GPU}{graphics processing unit}
\newacronym{gpgpu}{GPGPU}{general-purpose computing on graphics processing units}
\newacronym{fpga}{FPGA}{field-programmable gate array}
\newacronym[longplural={systems-on-chip}]{soc}{SoC}{system-on-a-chip}
\newacronym{dvfs}{DVFS}{dynamic voltage and frequency scaling}
\newacronym{pmc}{PMC}{performance monitoring counter}
\newacronym{isa}{ISA}{instruction set architecture}
\newacronym{pmu}{PMU}{performance monitoring unit}
\newacronym{psu}{PSU}{power supply unit}
\newacronym{sm}{SM}{streaming multiprocessor}
\newacronym{cupti}{CUPTI}{CUDA Performance Tools Interface}
\newacronym{gpuperfapi}{GPUPerfAPI}{GPU Performance API}
\newacronym{i2c}{I2C}{Inter Integrated Circuit}
\newacronym{adc}{ADC}{analog-to-digital converter}
\newacronym{lls}{LLS}{linear least squares}
\newacronym{nnls}{NNLS}{non-negative least squares}
\newacronym{pcc}{PCC}{Pearson correlation coefficient}
\newacronym{lut}{LUT}{lookup table}
\newacronym{bpu}{BPU}{branch prediction unit}
\newacronym{alu}{ALU}{arithmetic-logic unit}
\newacronym{simd}{SIMD}{single instruction, multiple data}
\newacronym{api}{API}{application programming interface}
\newacronym{dpm}{DPM}{dynamic power management}
\newacronym{cmos}{CMOS}{complementary metal–oxide–semiconductor}
\newacronym{soa}{SoA}{state of the art}
\newacronym{hpc}{HPC}{high performance computing}
\newacronym{mape}{MAPE}{mean absolute percentage error}
\newacronym{svr}{SVR}{support vector regression}
\newacronym{os}{OS}{operating system}
\newacronym{ann}{ANN}{artificial neural network}
\newacronym{mac}{MAC}{multiply–accumulate}
\newacronym{igpu}{iGPU}{integrated GPU}

\addto\extrasenglish{%
}
\addto\extrasenglish{%
}
\addto\extrasenglish{%
}

\begin{document}
\title{A Data-Driven Approach to Lightweight DVFS-Aware Counter-Based Power Modeling for Heterogeneous Platforms}
\titlerunning{A Data-Driven Approach to Counter-Based Power Modeling}
%
\author{\mbox{Sergio Mazzola\inst{1}\orcidID{0000-0001-8705-8990}} \and
\mbox{Thomas Benz\inst{1}\orcidID{0000-0002-0326-9676}} \and
\mbox{Björn Forsberg\inst{1}} \and
\mbox{Luca Benini\inst{1,2}\orcidID{0000-0001-8068-3806}}}
\authorrunning{S. Mazzola et al.}
\institute{Integrated Systems Laboratory (IIS), ETH Zürich, Zürich, Switzerland
\email{\{smazzola,tbenz,bjoernf,lbenini\}@iis.ee.ethz.ch} \and
DEI, University of Bologna, Bologna, Italy}
\maketitle
\begin{abstract}
Computing systems have shifted towards highly parallel and heterogeneous architectures to tackle the challenges imposed by limited power budgets.
These architectures must be supported by novel power management paradigms addressing the increasing design size, parallelism, and heterogeneity while ensuring high accuracy and low overhead.
In this work, we propose a systematic, automated, and architecture-agnostic approach to accurate and lightweight \acrshort{dvfs}-aware statistical power modeling of the CPU and GPU sub-systems of a heterogeneous platform, driven by the sub-systems' local \glspl{pmc}. Counter selection is guided by a generally applicable statistical method that identifies the minimal subsets of counters robustly correlating to power dissipation. Based on the selected counters, we train a set of lightweight, linear models characterizing each sub-system over a range of frequencies. Such models compose a lookup-table-based system-level model that efficiently captures the non-linearity of power consumption, showing desirable responsiveness and decomposability.
We validate the system-level model on real hardware by measuring the total energy consumption of an NVIDIA Jetson AGX Xavier platform over a set of benchmarks. The resulting average estimation error is 1.3\%, with a maximum of 3.1\%. Furthermore, the model shows a maximum evaluation runtime of 500\;ns, thus implying a negligible impact on system utilization and applicability to online \gls{dpm}.

\end{abstract}
%
%
%


\section{Introduction}\label{sec:intro}
Power concerns are shaping general-purpose computing in the post-Dennard scaling era, imposing a transistor utilization wall determined by the increasingly high power density of chips. The \textit{dark silicon} challenge triggered the shift from general-purpose processing units to highly parallel and specialized accelerators, able to achieve higher performance and energy efficiency at the price of increased architectural heterogeneity \cite{taylor2012dark}.

Power management paradigms addressing such architectures are a priority: today, dynamic hardware adaptations are extensively exploited to meet power and thermal constraints. \Gls{dvfs} and clock gating are two common examples of the \gls{dpm} techniques that allow tuning energy consumption to the application demand.
For power management to be possible, online power measurement is a requirement. Several off-the-shelf \glspl{soc} come equipped with built-in power sensors; however, such a solution is often problematic due to its final deployment costs, poor scalability, the coarse granularity of measurements, and low speed \cite{lin2020taxonomy}. A robust \gls{dpm} control loop, on the other hand, requires fast and accurate measurements, in addition to resource-level introspection for fine-grained power tuning.

It is well-known that \acrfullpl{pmc} activity correlates with the power consumption of diverse hardware components \cite{bellosa2000benefits,bircher2011complete,wang2019statistic}. As part of the digital domain, their usage is cheap, and their readings are fast and reliable, thus paving the way for accurate and efficient power models for \gls{dpm} policies.
\Glspl{pmc} also provide a high degree of introspection for the power consumption of individual components \cite{isci2003runtime}, which empowers \gls{pmc}-based models with desirable responsiveness and decomposability \cite{bertran2012systematic}.

Modern computer architectures expose hundreds of countable performance events \cite{nvidiaxavier,armcortexA57}. Hence, the parameter selection for a robust statistical power model often requires considerable knowledge of the underlying hardware, compromising the generality of this approach while increasing its complexity and deployment time. Growing parallelism and heterogeneity, together with the frequent lack of open documentation, further hinder this challenge.
A careful choice of the model parameters is necessary for several additional factors: 
first, \glspl{pmu} can simultaneously track only a limited number of performance counters. Also, the amount of model predictors directly impacts its evaluation overhead, which is required to be small for practical \gls{dpm} strategies and minimal interference with regular system operation. \Gls{dvfs} determines an additional layer of modeling complexity, as hardware behavior at varying frequencies has to be considered.

In this work, we develop a data-driven statistical approach to \gls{pmc}-based power modeling of modern, \gls{dvfs}-enabled heterogeneous computing systems. While targeting the multi-core CPU and the GPU of a heterogeneous platform, our methodology can be extended to include any additional sub-system, like memories and accelerators.
With respect to the extensive research carried out in the field of statistical \gls{pmc}-based power modeling, the distinctive trait of our approach is the \textit{combination} of general applicability, automated model construction, lightweight model evaluation, and high accuracy. In detail, our main contributions are:

\begin{itemize}
    \item a data-driven, automatic, and architecture-independent framework to select the \glspl{pmc} best representing the power consumption of the CPU and the GPU sub-systems of a generic target platform;
    \item the development of a system-level power model based on a \gls{lut}, composed of a set of lightweight, linear power models, one for each \gls{dvfs} frequency of each sub-system; this approach addresses the platform parallelism and heterogeneity while modeling \gls{dvfs} non-linearities with simple, low-overhead computation and limited modeling complexity;
    \item the training and the validation of the mentioned power models against real hardware: an off-the-shelf modern heterogeneous system, the NVIDIA Jetson AGX Xavier \cite{nvidiaxavier} board.
\end{itemize}

The rest of the paper is organized as follows: in \autoref{sec:rel_work} we depict the \gls{soa} to place our work into perspective; \autoref{sec:methodology} details our methodology and its background, while \autoref{sec:exp_setup} describes its implementation on an off-the-shelf heterogeneous modern platform. Finally, in \autoref{sec:results} we evaluate the results of our models' implementation, drawing conclusions in \autoref{sec:concl}.


\section{Related work}\label{sec:rel_work}

In the last 20 years, \gls{pmc}-based statistical power modeling aimed to online usage has been explored from several points of view \cite{ahmad2017survey,lin2020taxonomy}. Bertran et al. \cite{bertran2013counter} identify two families of models, based on their construction: \textit{bottom-up} and \textit{top-down}. In this section, we adhere to this taxonomy to present the \gls{soa} and compare it to our proposal, with particular reference to \autoref{tab:soa}.

\textit{Bottom-up} approaches rely on extensive knowledge of the underlying architecture to estimate the power consumption of individual hardware sub-units.
Isci et al. \cite{isci2003runtime} pioneers this field by breaking down the consumption of a single-core Pentium 4 CPU into \gls{pmc}-based power models for 22 of its main sub-units. Bertran et al.'s more systematic approach \cite{bertran2012systematic} refines this idea by extending it to a \gls{dvfs}-enabled multi-core CPU. Sub-units models are \textit{composed} to obtain accurate and \textit{responsive} higher-level models. However, such a complex model building requires heavy manual intervention and architectural awareness, jeopardizing model generality. The reported \gls{mape} for power estimations is 6\%, but their results highlight the strict dependency between bottom-up models' accuracy and architectural knowledge of the platform.

\textit{Top-down} approaches target simple, low-overhead, and generally applicable models. They commonly assume a linear correlation between generic activity metrics and power consumption.
Among the first attempts in the field, Bellosa \cite{bellosa2000benefits} models the power consumption of a Pentium II processor with a few manually selected \glspl{pmc}. Subsequent works \cite{pusukuri2009methodology,singh2009real} refine the idea with a more elaborate procedure for \gls{pmc} selection and multi-core support. Bircher and John \cite{bircher2011complete} are the first to go towards a thorough system-level power model, tackling the issue from a top-down perspective for each sub-system.
To the best of our knowledge, no past research studies a combination of accurate and lightweight models addressing \gls{dvfs} without requiring expert architectural knowledge.

\begin{table}[t]
    \centering
    \caption{Comparison of our work with the \gls{soa} in statistical power modeling}
    \label{tab:soa}
    \begin{tabular}{@{}Z{73pt}M{51.72pt}M{51.72pt}M{51.72pt}M{56.44pt}M{49pt}@{}}
        \toprule
        & Bertran et al. (2012) \cite{bertran2012systematic} & Walker et al. (2016) \cite{walker2016accurate} & Wang et al. (2019) \cite{wang2019statistic} & Mammeri et al. (2019) \cite{mammeri2019performance} & Our work \\
        \midrule
        \footnotesize\textbf{Heterogeneity}         & CPU only & \footnotesize CPU only                 & \footnotesize iGPU only & \footnotesize \checkmark & \checkmark \\ \midrule
        \footnotesize\textbf{Generality}            & low & \footnotesize ARM cores                  & \footnotesize no             & \footnotesize mobile     & \checkmark \\ \midrule
        \footnotesize\textbf{Automation}            & low & \checkmark                               & \footnotesize no            & \footnotesize low            & \checkmark \\ \midrule
        \footnotesize\textbf{Architecture-agnostic} & no & \footnotesize{no, ARM}                    & \footnotesize no            & \footnotesize average                   & \checkmark \\ \midrule
        \footnotesize\textbf{Lightweight model}     & no & \checkmark                                & \checkmark            & \footnotesize no             & \checkmark \\ \midrule
        \footnotesize\textbf{DVFS support}          & \checkmark & \checkmark                        & \footnotesize no            & \footnotesize no             & \checkmark \\ \midrule
        \footnotesize\textbf{Decomposability}       & \checkmark & no                                & \footnotesize no            & \footnotesize no             & \checkmark \\ \midrule
        \footnotesize\textbf{Accuracy (MAPE)} & \footnotesize power \mbox{6\%--20\%} & \footnotesize power \mbox{3\%--4\%} & \footnotesize \footnotesize power 3\% & \footnotesize power 4.5\%   & \footnotesize power 7.5\% energy 1.3\%\\
        \bottomrule
    \end{tabular}
\end{table}

More recent works have targeted CPU power modeling in mobile and embedded platforms. Walker et al. \cite{walker2016accurate} employ a systematic and statistically rigorous technique for \gls{pmc} selection and train power models for ARM A7 and A15 embedded processors. However, only one trained weight is used to predict the power consumption at any \gls{dvfs} state, which can be prone to overfitting. In contrast to their work, we target a broad range of platforms with more straightforward statistical methodologies and models, nevertheless carefully dealing with \gls{dvfs} and reaching comparable accuracy numbers.

Top-down power modeling approaches have also been applied to GPUs.
Wang et al. \cite{wang2019statistic} analyze the power consumption of an AMD \gls{igpu}, carefully studying its architecture and selecting the best \glspl{pmc} to build a linear power model. Redundant counters are discarded to reduce the model overhead, with power \gls{mape} below 3\%. However, the generated model is not generally applicable and requires expert knowledge.
Recent works resort to deep learning for creating accurate black-box power models: Mammeri et al. \cite{mammeri2019performance} train an \gls{ann} with several manually chosen CPU and GPU \glspl{pmc}. With an average power estimation error of 4.5\%, they report power estimates 3$\times$ more accurate than a corresponding linear model; however, the overhead of evaluating a neural network at runtime is not negligible, as it requires a number of \glspl{mac} two orders of magnitude higher than for a linear model. Potentially long training time, complex manual selection of the neural network topology, and lack of decomposability are additional drawbacks of this approach.

Our model shares its decomposability and responsiveness with bottom-up approaches but resorts to top-down modeling for individual sub-systems: we trade a lower per-component introspection for a systematic modeling procedure requiring very little architectural knowledge and minimal human intervention, targeting broad applicability. We indeed refine the approaches in \cite{pusukuri2009methodology,bircher2011complete,pi2019study} to allow the selection of a minimal set of best \glspl{pmc} for accurate and lightweight power models.
Our \gls{lut}-based approach addresses the platform heterogeneity and its \gls{dvfs} capabilities while employing lightweight linear models.
To the best of our knowledge, no previous work thoroughly combines all the mentioned features.
In this paper, we focus on power modeling for the CPU and GPU sub-systems of a heterogeneous platform, but our methodology can be extended to any additional sub-system, either considering its local counters, if available, or its related CPU counters as in \cite{bircher2011complete}.


\section{Methodology}\label{sec:methodology}

The main focus of our work is \gls{dvfs}-aware power consumption estimation of heterogeneous systems based on individual sub-systems modeling. To achieve this, we develop a \gls{lut}-based statistical power model of our target platform (\autoref{fig:teaser_scheme}, \Circled[]{\textbf{12}}). Two parameters index the \gls{lut}: the sub-system $d$ to be modeled and its current operational frequency $f_d$, corresponding to $d$'s \gls{dvfs} state. Each entry of the \gls{lut} is a linear power model $P_{d}$. The model is driven by $X_{d, f_{d}}$, the set of $d$'s \glspl{pmc} tracked when operating at $f_d$, weighted by the set of trained weights $W_{d, f_{d}}$:

\begin{figure}[t]
  \centering
  \includegraphics[width=\textwidth,height=5cm]{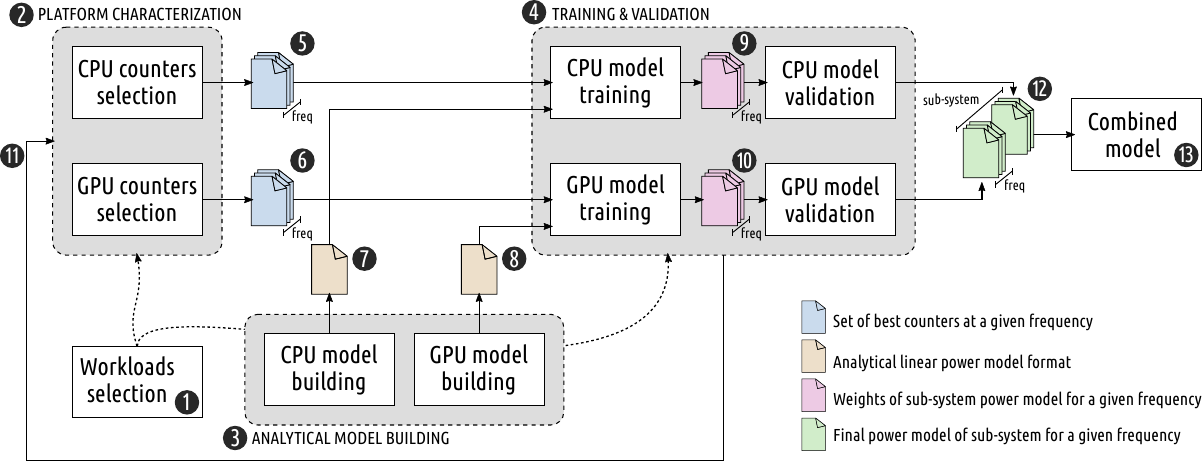}
  \caption{Scheme of the proposed data-driven, automatic power modeling approach for DVFS-enabled heterogeneous platforms}
  \label{fig:teaser_scheme}
\end{figure}

\begin{equation}\label{eq:lut}
    LUT[d][f_{d}] = P_{d}(X_{d, f_{d}}, W_{d, f_{d}}),\quad d \in D^\ast\ \textrm{and}\ f_d \in F_d^\ast
\end{equation}

We refer to the process of defining $X_{d, f_{d}}$ for each $d$ at each $f_d$ as \textit{platform characterization} (\autoref{fig:teaser_scheme}, \Circled[]{\textbf{2}}). It is architecture-agnostic and performed one time per target system.
We refer to the set of all sub-systems of the platform as $D$. For each $d \in D$, its available \gls{dvfs} states are denoted by the set $F_d$. Our methodology targets a subset of sub-systems $D^\ast \subseteq D$, and, for each $d \in D^\ast$, a subset of frequencies $F_d^\ast \subseteq F_d$. Both $D^\ast$ and $F_d^\ast$ are user-defined parameters. For our methodology, we consider $D^\ast = \{CPU,\ GPU\}$. However, the described approach can be extended to any $D^\ast \subseteq D$. The choice of $F_d^\ast$ is discussed in \autoref{subsec:methdology_impl}.

The power models $P_{d}$ for the two sub-systems are built independently (\autoref{fig:teaser_scheme}, \Circled[]{\textbf{3}} and \Circled[]{\textbf{4}}); their composition results in an accurate and decomposable system-level power model (\autoref{fig:teaser_scheme}, \Circled[]{\textbf{13}}), developed with minimal effort.
For the whole procedure, we operate a careful choice in terms of workloads for a representative dataset (\autoref{fig:teaser_scheme}, \Circled[]{\textbf{1}}).
With reference to \autoref{fig:teaser_scheme}, we describe the several steps in the following.
    
    \subsection{Analytical model building}\label{subsec:power_models_build}
    
    For each sub-system $d$, we individually develop an analytical power model (\autoref{fig:teaser_scheme}, \Circled[]{\textbf{3}}). As the frequency $f_d$ is factored out by the \gls{lut}, the individual $P_{d}$ (\autoref{fig:teaser_scheme}, \Circled[]{\textbf{7}} and \Circled[]{\textbf{8}}) are linear models; however, their \gls{lut} composition is still able to capture the non-linearities of \gls{dvfs} state variations.
    The analytical formats of $P_{CPU}$ and $P_{GPU}$ are discussed in the following. 
    
        \subsubsection{CPU sub-system power model}\label{subsubsec:cpu_model_build}
        
            Our analytical CPU power model is based on the classical \gls{cmos} power model, and assumes per-cluster \gls{dvfs}: $P_{CPU} = P_{leak} + P_{dyn} = P_{leak} + \alpha C V^2 f$. For the sake of model simplicity, we consider the static power to be equal to the leakage $P_{leak}$.
            To account for possibly hardware-managed power gating, we include it in our model by defining the metric $g_i$, i.e., the percentage of time the core $i$ is not power-gated.
            The dynamic power $P_{dyn}$ of the whole CPU can then be defined as the summation of each core $i$'s contribution $P_{dyn, i}$ over all cores as
            \begin{equation}\label{eq:cpu_dyn_power}
                P_{dyn} = \sum_{i = 1}^{\#cores} \Big[ (1-g_i)\cdot \Big( P_{gate, i} - P_{leak, i} \Big) +  \alpha_i C_i V^2 f \Big]
            \end{equation}
            where $P_{gate, i} - P_{leak, i} < 0$ and accounts for the leakage power saved when core $i$ is gated, while $\alpha_i C_i V^2 f$ represents its activity when non-gated. $P_{gate, i}$ and $P_{leak, i}$ are respectively the gated power and the leakage power of core $i$.
            The term $g_i$ is computed using each core's cycle counter, while we resort to the CPU performance counters $x_{ij} \in X_{CPU, f}$ coming from the platform characterization to model the switching activity $\alpha_i$.
            Factoring out of the summation $P_{gate}$, which represents the power dissipated when all cores are power-gated, we can highlight the contribution of each core $i$ and each \gls{pmc} $x_{ij}$ to $P_{CPU}$ rewriting it as
            \begin{equation}\label{eq:cpu_power_2}
                P_{CPU} = P_{gate} + \sum_{i = 1}^{\#cores} \Big[ 
                    g_i\cdot \Big( P_{leak, i} - P_{gate, i} \Big) +
                    \sum_{j = 1}^{\#PMCs} x_{ij} C_i V^2 f
                \Big]
            \end{equation}
            The core parasitic capacitance $C_i$ is a constant of the platform; additionally, as $d = CPU$ and $f_d = f$ index the \gls{lut} from \autoref{eq:lut}, $V$ and $f$ are also constant: consequently, we lump them into model weights. We do the same for \mbox{$P_{leak, i} - P_{gate, i}$}, which is now positive. The resulting analytical expression, in \autoref{eq:cpu_power_model}, can be trained with a linear regression.
            \begin{equation}\label{eq:cpu_power_model}
                P_{CPU} = L + \sum_{i = 1}^{\#cores} \Big( g_i\cdot G_i + 
                    \sum_{j = i}^{\#PMCs} x_{ij}\cdot A_{ij} \Big)
            \end{equation}
            $L$, $G_i$ and $A_{ij}$ are the parameters of the model, from  the $W_{CPU, f}$ set. $g_i$ and $x_{ij}$ are the input independent variables, or the \textit{predictors}.
        
        \subsubsection{GPU sub-system power model}
        
            Like for the CPU, the GPU power model (\autoref{fig:teaser_scheme}, \Circled[]{\textbf{8}}) also considers leakage and dynamic power. The same considerations of the CPU model hold as the \gls{lut} approach fixes the \gls{dvfs} state for $P_{GPU}$.
            However, due to commonly stricter limitations of GPU's \glspl{pmu} (cf. \autoref{subsec:platform_char}), it would not be general enough to always assume the availability of a clock cycle counter.
            Consequently, the model $P_{GPU}$ is simply driven by the \glspl{pmc} $x_{j} \in X_{GPU, f}$ coming from the platform characterization.
            \begin{equation}\label{eq:gpu_power_model}
                P_{GPU} = K + \sum_{j = i}^{\#PMCs} x_{j}\cdot B_{j}
            \end{equation}
            The model parameters in \autoref{eq:gpu_power_model} are $K$, modeling the constant leakage power, and $B_j$, weighting the dynamic activity. Before feeding the model, the counters $x_j$ are normalized to the sampling period: as no other measures of time are present in $P_{GPU}$, this is required for robustness to sampling jitter.
        
    \subsection{Platform characterization}\label{subsec:platform_char}
    
        Independently for each sub-system $d$ and frequency $f_d$, we perform a one-time correlation analysis between all of its local \glspl{pmc} and the sub-system power consumption, looking for the optimal $X_{d, f_{d}}$ set in terms of model accuracy, low overhead and compliance with \gls{pmu} limitations (\autoref{fig:teaser_scheme}, \Circled[]{\textbf{5}} and \Circled[]{\textbf{6}}). We refer to this procedure as \textit{platform characterization}.
        Our characterization is meant as an automatic, architecture-agnostic, and data-driven alternative to manual intervention when it comes to \gls{pmc} selection.
        Given the sub-system $d$, its characterization involves the following steps:
    
        \begin{enumerate}
            \item for each \gls{dvfs} state $f_d \in F_d^\ast$, profile \textit{all} performance events exposed by $d$ while tracing $d$'s power; to to track all events despite the \gls{pmu} limitations, we make use of \textit{multiple passes} \cite{amdgpuperfapi,nvidiacupti} and avoid \textit{counter multiplexing} \cite{may2001mpx}\textsuperscript{\ref{fn:counter_mux}};
            \footnotetext[1]{\label{fn:counter_mux}
                With \textit{counter multiplexing} the \glspl{pmc} are time-multiplexed in a round-robin fashion; this increases overhead and decreases accuracy. To avoid it when profiling \textit{all} available events, we perform \textit{multiple passes}: given a set of performance events that cannot be thoroughly profiled simultaneously, we replay the same workload (i.e., we induce the same power consumption profile) until all events have been tracked.
            }
            \item normalize the \gls{pmc} samples with respect to the sampling periods to work around sampling jitter;
            \item compute a \gls{lls} regression of each \gls{pmc}'s activity trace over its related power measurements, for each $f_d$; events with a p-value above 0.05 are discarded as not reliable for a linear correlation; 
            \item individually for each $f_d$, sort the remaining events by their \gls{pcc} and select the best ones that can be profiled simultaneously (i.e., compose $X_{d, f_{d}}$ for all $f_d \in F_d^\ast$).
        \end{enumerate}%
        The generated $X_{d, f_{d}}$ counter sets (\autoref{fig:teaser_scheme}, \Circled[]{\textbf{5}} and \Circled[]{\textbf{6}}) are directly employed to build the linear models for each $d$ and $f_d$ in the \gls{lut} (\autoref{eq:lut}).
        While multiple passes are used to profile all available events during the offline platform characterization, with an online deployment in mind for our models, we require $X_{d, f_{d}}$ to only contain events that can be tracked simultaneously by the \gls{pmu}.
        To this end it is usually enough to select, for each frequency $f_d$, the desired number of best counters, up to the \gls{pmu} limit.
        To find the optimal number of performance counters with respect to model accuracy, results from model evaluation can be employed (\autoref{fig:teaser_scheme}, arrow \Circled[]{\textbf{11}}).
        
        On the other hand, some platforms \cite{nvidiacupti,amdgpuperfapi,may2001mpx} might have stricter \gls{pmu} constraints. Not only \glspl{pmc} are limited in number, but some of them are mutually exclusive: \textit{compatibility} must also be considered. As documentation about performance events compatibility is not usually available, we tackle this issue with a \gls{pmu}-aware iterative algorithm based on the profiling \gls{api} provided by the vendor. 
        Given the sub-system $d$, its frequency $f_d$ and the list of \glspl{pmc}, the algorithm heuristically tries to group the \glspl{pmc} with highest \gls{pcc} one by one, adding to $X_{d, f_{d}}$ only events that, basing on the provided \gls{api}, can be counted in a single pass.
    
    \subsection{Train, Validate, Combine}\label{subsec:train_valid}
    
        With the sets of counters $X_{d, f_{d}}$ from platform characterization we compose the \gls{lut} from \autoref{eq:lut} by individually training a linear power model $P_{d}$ for each sub-system $d \in D^\ast$ at each $f_d \in F_d^\ast$ (\autoref{fig:teaser_scheme}, \Circled[]{\textbf{4}}). The output of each training is a set of weights $W_{d, f_d}$ (\autoref{fig:teaser_scheme}, \Circled[]{\textbf{9}} and \Circled[]{\textbf{10}}).
        We then validate each individual $P_{d}(X_{d, f_{d}}, W_{d, f_{d}})$. The final result of this process is the complete \gls{lut} to model the power of all the components in the heterogeneous system (\autoref{fig:teaser_scheme}, \Circled[]{\textbf{12}}).
        
        For this step, we build a dataset with the characteristics discussed in \autoref{subsec:workload_sel} by extracting traces of workloads activity and power consumption.
        We acquire data points in a continuous, periodical sampling mode to collect information at a fine grain such that the desired responsiveness, necessary for \gls{dpm}, can be achieved by the resulting model \cite{bertran2012systematic}.
        To train each individual $P_{d}(X_{d, f_{d}}, W_{d, f_{d}})$, we perform a \gls{nnls} linear regression of the \glspl{pmc} values over the power measurements, obtaining the set of non-negative weights $W_{d, f_{d}}$. Compared to unconstrained \gls{lls}, non-negative weights are physically meaningful, and prove to be robust to multicollinearity, which makes them less prone to overfitting.
    
        After individual training and validation, we combine the CPU and GPU models into a system-level power model (\autoref{fig:teaser_scheme}, \Circled[]{\textbf{13}}) defined as $P_{SYS} = \sum_{d\in D^\ast} P_{d}(X_{d, f_{d}}, W_{d, f_{d}})$. In other words, the system-level power model is the \textit{reduction sum} of the \gls{lut} in \autoref{eq:lut} along the sub-systems dimension $d$.
        Modeling the sub-systems individually until this step allows us to avoid profiling all possible pairs of $\langle f_{CPU},\ f_{GPU} \rangle$ for training, which is simpler, faster, and more robust to overfitting. The final model is decomposable, accurate, and computationally lightweight.
        We validate the combined model with a dataset built by concurrently profiling activity and power consumption of CPU and GPU.

    \subsection{Workloads selection}\label{subsec:workload_sel}
    
    The platform characterization, training, and validation steps are based on the careful choice of a representative set of workloads for the heterogeneous platform.
    Our workloads selection (\autoref{fig:teaser_scheme}, \Circled[]{\textbf{1}}) is mainly driven by the following parameters:
    firstly, complete coverage of all targeted sub-systems is required to fully address the heterogeneity of the platform.
    Secondly, for each sub-system, the workload should be able to induce a broad range of different behaviors for a workload-independent result.
    
    Our reference for this is Rodinia 3.1 \cite{che2009rodinia}, a benchmark suite designed for heterogeneous and parallel systems supporting OpenMP, OpenCL, and CUDA. For the CPU characterization and modeling, we employ Rodinia's OpenMP benchmarks; for the GPU, we use the CUDA benchmarks.
    For the composition of the Rodinia suite, Che et al. follow Berkeley's dwarf taxonomy \cite{asanovic2006landscape}, which fulfills the variety of behaviors we require.
    Furthermore, we increase such variety by developing additional synthetic benchmarks targeting generic CPU sub-units to minimize the possibility of biasing the model towards the training set \cite{bertran2012systematic,bertran2013counter}.
    Such insights have not been reported for GPU power models.


\section{Experimental setup}\label{sec:exp_setup}

For the implementation of our methodology, we target a modern \gls{dvfs}-enabled heterogeneous system, the NVIDIA Jetson AGX Xavier.
In this section, we describe our measurement environment and data acquisition setup, putting everything together for the implementation and practical demonstration.

    \subsection{Target platform description}
    
        The target platform for our experiments is an NVIDIA Jetson AGX Xavier, powered by the Xavier
        \gls{soc} \cite{nvidiaxavier}. It is a highly parallel and heterogeneous \gls{soc} provided with an 8-core 64-bit ARMv8.2 CPU, a 512-core NVIDIA Volta GPU and several additional accelerators for deep-learning, computer vision, and video encoding/decoding.
        With several \gls{dvfs} \emph{power profiles} available for its sub-systems, this platform represents a fruitful target to explore the issues tackled by this paper. In particular, the CPU complex can be clocked at 29 different discrete frequencies between 115\;MHz and 2266\;MHz, while the GPU has 14 available \gls{dvfs} states between 115\;MHz and 1377\;MHz.

    \subsection{Data acquisition}\label{subsec:data_acq}
    
        \subsubsection{Power measurements}\label{subsubsec:power_measures}
    
            The NVIDIA board features two 3-channel INA3221 power monitors whose information can be read using \gls{i2c} sysfs nodes. In particular, they can measure the power consumption of the CPU and GPU power domains independently.
            
            As mentioned in \autoref{sec:intro}, on-board power monitors are not robust tools for \gls{dpm} mainly due to their speed, in this case limited by the \gls{i2c} protocol, coarse time granularity, due to their analog nature, and low resolution, which for the Xavier is limited to approximately 200\;mW due to the board configuration.
            However, their usage does not require external equipment, and they can be programmatically and reliably driven. Hence, they are still helpful for building datasets to achieve higher introspection, time granularity, and responsiveness enabled by \glspl{pmc}-based power models.
    
        \subsubsection{Activity measurements}\label{subsec:activity_measure}
            Based on an ARMv8 \gls{isa}, the NVIDIA Carmel CPU Complex exposes 84 performance events per core \cite{armcortexA57}. The \gls{pmu} of each core implements three 32-bit hardware performance counters, which can track any of the available events; an additional 64-bit counter dedicated to clock cycles is also present.
            This particularly suits the CPU power model of \autoref{eq:cpu_power_model}.
            To keep a low profiler overhead, we avoid higher-level tools such as perf. Instead, we directly access the \gls{pmu} counters by reading their memory-mapped registers via assembly instructions.
        
            The NVIDIA Volta GPU of the Xavier \gls{soc} features eight 64-core \glspl{sm}. Its \gls{pmu} exposes a total of 61 performance events: 41 of them are instantiated in each of the eight \glspl{sm}, while the remaining 20 are related to the L2 cache and have four instances each. The maximum number of events that the \gls{pmu} can track in parallel is not fixed and depends on their compatibility to be counted simultaneously.
            NVIDIA provides developers with several tools to retrieve GPU \gls{pmc} information \cite{pi2019study}; we employ \gls{cupti} \cite{nvidiacupti}, which features the lowest reading overhead.
    
    \subsection{Methodology implementation}\label{subsec:methdology_impl}
        
        With reference to the notation of \autoref{sec:methodology}, we implement our methodology on an NVIDIA Jetson AGX Xavier platform targeting a sub-system set $D^\ast = \{CPU,\ GPU\}$ and the following \gls{dvfs} states sets:
        \begin{itemize}
            \item $F_{CPU}^\ast = \{730,\ 1190,\ 2266 \}$;
            \item $F_{GPU}^\ast = F_{GPU} = \{ \text{all 14 frequencies from 115 to 1377} \}$.
        \end{itemize}
        \footnotetext[2]{\label{fn:freq_rounding}
            We round decimal frequency values to integers for readability.
        }
        All $f_d$ frequencies are expressed in MHz\textsuperscript{\ref{fn:freq_rounding}}.
        Note that, at lower CPU operating frequencies, \gls{os} interference becomes predominant, affecting our measurements. Consequently, we limit our study to a subset of the most generally useful frequencies for the host CPU.
        
        Profiling a workload on the target platform is necessary during several steps of our methodology. To this end, we developed a profiler that entirely runs on the Xavier board, tracing the \gls{pmc} of each sub-system while collecting power measures, as outlined in \autoref{subsec:data_acq}. Samples are acquired in a continuous mode with a sampling period of 100\;ms. Higher sampling periods turn our not to gain additional information due to the electrical inertia of the on-board analog power monitors.
        The profiler samples \glspl{pmc} and power values in the same sampling
        %
\noindent
\begin{minipage}{\textwidth}
    \begin{minipage}[b]{0.42\textwidth}
        \centering
        \captionof{table}{The ten best CPU counters from platform characterization at each frequency, with their \gls{pcc}. The counters selected for the model are reported in green.}
        \label{tab:cpu_corr_analysis}
        \begingroup
        \renewcommand{\arraystretch}{1.013} 
        \begin{tabular}[b]{@{}Z{2.3cm}ccc@{}}
            \toprule
            \multirow{2}{*}{\begin{tabular}[c]{@{}l@{}}Performance\\event\end{tabular}} & \multicolumn{3}{c}{Frequency [MHz]} \\
             & \footnotesize 730 & \footnotesize 1190 & \footnotesize 2266 \\
            \midrule
            \footnotesize{Cycles counter} &           \textbf{\small{0.56}} &  \textbf{\small{0.57}}       &  \textbf{\small{0.60}}  \\
            \footnotesize{Exception taken} &              \small{0.51} &  \small{0.54}       &   \color{mygreen}\textbf{\small{0.57}}  \\
            \footnotesize{Instr. retired} &         \color{mygreen}\textbf{\small{0.52}} &  \color{mygreen}\textbf{\small{0.57}}        &  \small{0.55}  \\
            \footnotesize{FP activity} &                  \color{mygreen}\textbf{\small{0.54}} &  \color{mygreen}\textbf{\small{0.56}}        &  \color{mygreen}\textbf{\small{0.59}}  \\
            \footnotesize{SIMD activity} &                  \small{0.50} &    \small{0.52}  &   \small{0.52}  \\
            \footnotesize{Speculative branch} &              \small{0.49} &  \small{0.52}        &  \small{0.53}  \\
            \footnotesize{Speculative load} &              \small{0.52} &  \small{0.53}       &   \small{0.54}  \\
            \footnotesize{Speculative L/S} &            \small{0.51} &  \small{0.53}        &  \small{0.53}  \\
            \footnotesize{L1 I\$ access} &            \color{mygreen}\textbf{\small{0.53}} &  \color{mygreen}\textbf{\small{0.54}}        &  \color{mygreen}\textbf{\small{0.56}}  \\
            \footnotesize{L1 D\$ access} &            \color{gray}\small{n/d} &  \small{0.54}        &   \color{gray}\small{n/d} \\
            \footnotesize{Data memory access (read)} &       \small{0.52} &  \small{0.53}       &   \small{0.54}  \\
            
            \bottomrule
        \end{tabular}
    \endgroup
    \end{minipage}
    \hfill
    \begin{minipage}[b]{0.55\textwidth}
        \centering
        \captionof{table}{The ten GPU \glspl{pmc} selected by our \gls{pmu}-aware algorithm at each frequency based on platform characterization; their \gls{pcc} with GPU power consumption is reported. The finally selected counters are reported in green.}
        \label{tab:smart_counters_gpu}
        \begin{tabular}[b]{@{}rZ{3.48cm}ccc@{}}
                \toprule
                \multicolumn{2}{c}{Performance event} & \multicolumn{3}{c}{Frequency [MHz]} \\
                \# & Description & \footnotesize 115 & \footnotesize 829 & \footnotesize 1377 \\
                \midrule
                \footnotesize{1}  & \footnotesize{CTAs launched} & \color{gray}\small{n/d.} & \color{gray}\small{n/d.} & \small{0.45} \\
                \footnotesize{2}  & \footnotesize{Cycles active warp (SM)} & \color{mygreen}\textbf{\small{0.77}} & \color{mygreen}\textbf{\small{0.67}} & \color{mygreen}\textbf{\small{0.65}} \\
                \footnotesize{3}  & \footnotesize{Acc. warps/cycle (SM)}  & \color{mygreen}\textbf{\small{0.74}} & \color{mygreen}\textbf{\small{0.65}} & \color{mygreen}\textbf{\small{0.57}} \\
                \footnotesize{9}  & \footnotesize{Cycles w/ active warps}     & \color{mygreen}\textbf{\small{0.58}} & \small{0.48} & \small{0.46} \\
                \footnotesize{10} & \footnotesize{Acc. warps per cycle}      & \color{mygreen}\textbf{\small{0.55}} & \small{0.47} & \color{gray}\small{n/d.} \\
                \footnotesize{19} & \footnotesize{Cycles w/o issued instr.}      & \color{mygreen}\textbf{\small{0.55}} & \color{gray}\small{n/d.} & \color{mygreen}\textbf{\small{0.55}} \\
                \footnotesize{20} & \footnotesize{Cycles w/ 1 issued instr.}      & \small{0.54} & \color{mygreen}\textbf{\small{0.53}} & \color{mygreen}\textbf{\small{0.55}} \\
                \footnotesize{21} & \footnotesize{Instr. executed per warp}     & \color{mygreen}\textbf{\small{0.54}} & \color{mygreen}\textbf{\small{0.54}} & \color{mygreen}\textbf{\small{0.55}} \\
                \footnotesize{22} & \footnotesize{Active threads instr.}  & \small{0.53} & \color{gray}\small{n/d.} & \color{gray}\small{n/d.} \\
                \footnotesize{23} & \footnotesize{Active and not pred. off threads instr.}   & \color{gray}\small{n/d.} & \color{mygreen}\textbf{\small{0.55}} & \color{mygreen}\textbf{\small{0.55}} \\
                \footnotesize{29} & \footnotesize{LDG instr. executed}       & \color{gray}\small{n/d.} & \color{mygreen}\textbf{\small{0.54}} & \color{gray}\small{n/d.} \\
                \footnotesize{45} & \footnotesize{Reads T\$ to L2, slice 0} & \color{mygreen}\textbf{\small{0.72}} & \color{mygreen}\textbf{\small{0.76}} & \color{mygreen}\textbf{\small{0.76}} \\
                \footnotesize{53} & \footnotesize{Reads T\$ to L2, slice 1}   & \color{mygreen}\textbf{\small{0.72}} & \color{mygreen}\textbf{\small{0.76}} & \color{mygreen}\textbf{\small{0.75}} \\
                \bottomrule
        \end{tabular}
    \end{minipage}
    \vspace{20pt}
\end{minipage}
        period, to grant the time correlation \cite{malony2011parallel} needed for an effective correlation analysis and training (cf. \autoref{subsec:platform_char} and \autoref{subsec:train_valid}).
        
        In terms of selected CPU workloads, we employ 17 different OpenMP benchmarks from the Rodinia suite in several multi-thread configurations and five additional synthetic benchmarks. For the GPU, we employ ten different CUDA benchmarks from Rodinia.
        To average out possible interference in our measurements, like unpredictable \gls{os} activity on the CPU, each workload is always profiled three times.


\section{Results}\label{sec:results}
    
    This section reports the result of our methodology applied to the NVIDIA Jetson AGX Xavier; we first go through the results of the \gls{pmc} selection, subsequently discussing the individual power models. We then show how the composed system-level power model achieves the objectives discussed in \autoref{sec:intro}.
    
    \begin{figure}[h!]
        \centering
        \begin{subfigure}[b]{0.49\textwidth}
            \centering
            \includegraphics{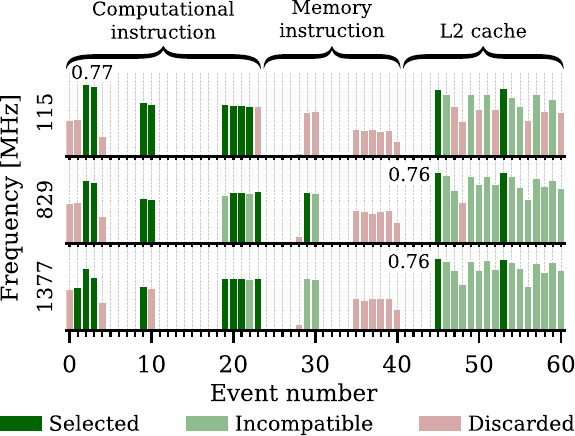}
            \caption{The \gls{pcc} of each GPU event at three frequencies, computed during platform characterization. The highest \gls{pcc} is annotated for each frequency. The different colors identify the result of our \gls{pmu}-aware selection algorithm.}
            \label{fig:gpu_corr_analysis}
        \end{subfigure}
        \hfill
        \begin{subfigure}[b]{0.49\textwidth}
            \centering
            \includegraphics{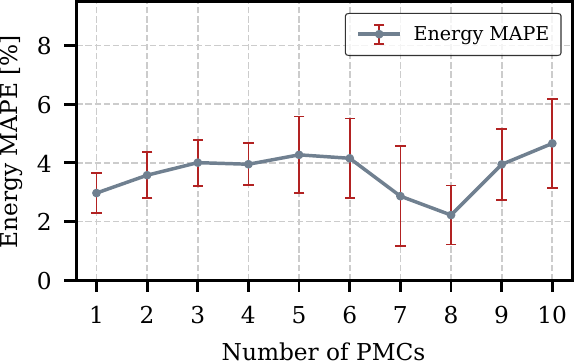}
            \caption{GPU power model energy estimation error as a function of the number of model predictors. The energy is accounted for the whole execution of the validation set, and its error is averaged over the estimations at all 14 frequencies.}
            \label{fig:gpu_counters_num}
        \end{subfigure}
        \caption{Results from the analysis for GPU \gls{pmc} selection.}
        \label{fig:gpu_study}
    \end{figure}

    \subsection{Platform characterization}\label{subsec:counters_sel_res}
    
        \subsubsection{CPU \gls{pmc} selection}\label{subsubsec:cpu_cnt_sel}
    
            For the CPU characterization, we profile the workloads reported in \autoref{subsec:methdology_impl}. \autoref{tab:cpu_corr_analysis} shows the main results for the CPU counter selection\textsuperscript{\ref{fn:counters_doc}}. For readability, only the ten best counters are shown, and the ones selected to model the power at each $f_d$ are highlighted. The \gls{pcc} of the clock cycle counter is also reported.
            
            \footnotetext[3]{\label{fn:counters_doc}
                Due to space limitations we abbreviate \gls{pmc} names in the tables; for a thorough description see their related documentation \cite{nvidiacupti,armcortexA57}.
            }
            
            Overall, the power consumption of the cores is highly correlated with the number of cycles of activity, the number of retired instructions, the floating-point activity, and various cache-related events.
            As no events compatibility issues emerged for the CPU, according to the \gls{pmu} capabilities, we consider the three best counters at each frequency and the additional \gls{pmu} register dedicated to the cycle counter.
            As the eight Carmel cores in the \gls{soc} are identical and only per-cluster \gls{dvfs} is supported by the platform, we choose to use the same \gls{pmc} configuration for all cores.
        
        \subsubsection{GPU \gls{pmc} selection}\label{subsubsec:gpu_cnt_sel}
                The results for the GPU characterization, obtained profiling the workloads discussed in \autoref{subsec:methdology_impl}, are shown in \autoref{fig:gpu_corr_analysis}; for readability, only three of the 14 explored frequencies are reported. The histograms report the \gls{pcc} of all 61 GPU events over the \gls{dvfs} states. As it turns out, Volta GPU's \gls{pmu} presents \gls{pmc} compatibility constraints (cf. \autoref{subsec:platform_char}); three event categories hence result from the execution of our \gls{pmu}-aware algorithm: \textit{selected}, the ones selected to drive the power model; \textit{incompatible}, the \glspl{pmc} with a high enough \gls{pcc} but discarded due incompatibility with better \glspl{pmc}; \textit{discarded}, all the remaining counters after the selection is completed.

                Our correlation analysis shows that the utilization rate of the L2 cache, a 512\;KB memory pool shared among all the GPU \glspl{sm}, is generally a good proxy for GPU power consumption, with most high-\gls{pcc} counters belonging to its domain.
                Nevertheless, from \autoref{fig:gpu_corr_analysis} it is clear how our selection algorithm grants enough \glspl{pmc} heterogeneity such that a broad range of GPU behaviors is captured: since a limited number of events can be usually tracked for the same domain (e.g., computational resources, L2 cache), the algorithm intrinsically selects non-redundant \glspl{pmc}, covering all GPU components and decreasing \textit{multicollinearity} among model predictors.
                
                Moreover, the \gls{pmu}-aware algorithm proves to be able to select at most ten performance events to be counted simultaneously, reported in \autoref{tab:smart_counters_gpu}\textsuperscript{\ref{fn:counters_doc}}; the table uses the same event numbering of \autoref{fig:gpu_corr_analysis}. Among those, we pick the optimal configuration by analyzing the energy estimation accuracy of the GPU power model as a function of the number of counters driving it (\autoref{fig:gpu_counters_num}). We opt for eight input variables: this implements the feedback from model evaluation discussed in \autoref{subsec:platform_char}.

    \begin{figure}[t]
        \centering
        \begin{subfigure}[b]{0.51\textwidth}
            \centering
            \includegraphics{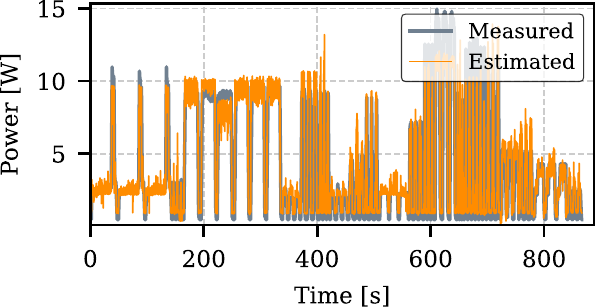}
            \caption{CPU sub-system, $f_{CPU}=$ 2266\;MHz}
            \label{fig:cpu_model_power}
        \end{subfigure}
        \hfill
        \begin{subfigure}[b]{0.48\textwidth}
            \centering
            \includegraphics{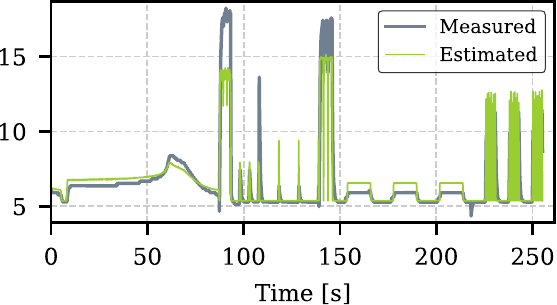}
            \caption{GPU sub-system, $f_{GPU}=$ 1377\;MHz}
            \label{fig:gpu_model_power}
        \end{subfigure}
        \caption{Example results from individual CPU and GPU power models validation (instantaneous power model estimates and power measures from INA3221)}
        \label{fig:power_models}
    \end{figure}
    
    \subsection{Sub-system models evaluation}\label{subsec:models_res}
    
        \subsubsection{CPU power model}\label{subsubsec:cpu_model_res}
        
            For the CPU, we train the linear model from \autoref{eq:cpu_power_model} with a \gls{nnls} regression for each frequency with a total of four independent variables per core.
            With the dataset discussed in \autoref{subsubsec:cpu_cnt_sel}, we use the initial 70\% for training and the remaining 30\% for validation.

            In terms of instantaneous power accuracy, the model achieves a \gls{mape} between 3\% and 4.4\% based on the frequency, with a standard deviation of approximately 5\%.
            
            As can be seen in \autoref{fig:cpu_model_power}, our model tracks the measured instantaneous power consumption of the CPU over time.
            The instantaneous behavior of the estimated power consumption is more volatile than the measured one. As a matter of fact, \glspl{pmc} are much more closely coupled to the workload and more responsive when compared to the INA3221 power monitors, which present the limitation discussed in \autoref{subsubsec:power_measures}.
            Our focus is on minimizing the error when integrating power over time, that is, calculating energy quantities. With a maximum energy estimation error of 4\%, which drops below 3\% at 1190\;MHz, our model delivers an accuracy competing with \gls{soa} or even superior.
        
        \subsubsection{GPU power model}\label{subsubsec:gpu_model_res}
        
            For the GPU, we likewise train the linear model of \autoref{eq:gpu_power_model} for each of the 14 GPU frequencies with a \gls{nnls} linear regression.
            The workload comprises the benchmarks discussed in \autoref{subsubsec:gpu_cnt_sel}; its initial 60\% is used as the training set, while the remaining 40\% is the validation set.
            
            As for the CPU case, we can effectively track the actual instantaneous power measured by the INA3221 monitors while obtaining higher time granularity (\autoref{fig:gpu_model_power}), achieving a power \gls{mape} between 6\% and 8\% (based on the frequency) in the instantaneous measurements. The standard deviation over all frequencies is approximately 8\%.
            When employed to estimate the energy over the full validation set, our model achieves a maximum error of 5.5\% over all frequencies, with an average of 2.2\%.
        
    \begin{figure}[t]
        \centering
        \includegraphics{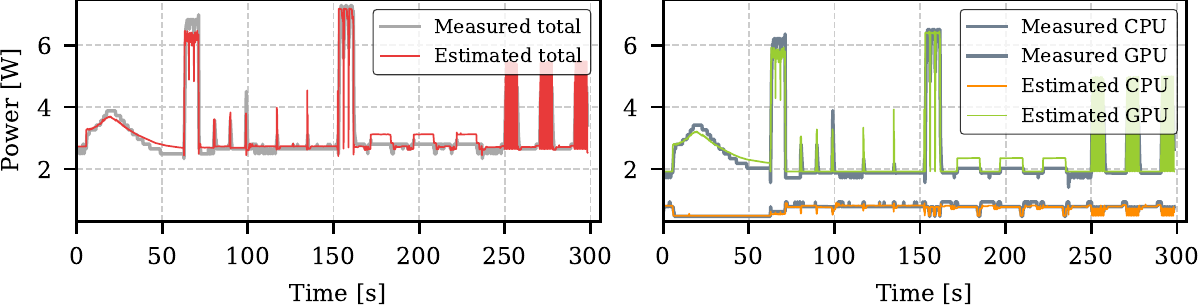}
        \caption{Instantaneous power estimate for the system-level power model and its power breakdown with $f_{CPU}=$ 1190\;MHz, $f_{GPU}=$ 829\;MHz}
        \label{fig:comb_model_power}
    \end{figure}
    
    \subsection{Combined model evaluation}\label{subsec:comb_model_res}
    
        After building, training, and validating both CPU and GPU power models individually, we put them together to obtain a system-level power model for every possible combination of $f_{CPU}$ and $f_{GPU}$.
        For the validation of the final, combined model, we employ the CUDA benchmark suite from Rodinia 3.1, as it proved to induce valuable activity in both of our sub-systems of interest.
        
        An example of the instantaneous power estimate is shown in \autoref{fig:comb_model_power}; the achieved instantaneous power \gls{mape} has an average of 8.6\% over all CPU and GPU frequency combinations.
        In terms of energy, it reaches an average estimation error of 2.5\%. A sub-sample of frequency combinations is reported in \autoref{tab:comb_energy} as an overview.
        However, our results highlight how the energy estimation error of the combined model is generally higher when $f_{CPU}$ and $f_{GPU}$ diverge from each other. In particular, when $f_{GPU}$ is very low compared to $f_{CPU}$, stalls and speculative behaviors may occur in the host CPU, which are hardly grasped by our model and jeopardize CPU's power estimation accuracy. As such configurations are generally not used in practice due to their sub-optimal efficiency, we focus on frequencies $f_{GPU} >$ 600\;MHz.
        With these considerations, we report an instantaneous power \gls{mape} of 7.5\% and energy estimation error of 1.3\%, with a maximum of approximately 3.1\%.
        To estimate the evaluation overhead of our system-level model, we implement it as a binary for the Carmel CPU and measure its runtime. Based on the frequency of the core, running between 730\;MHz and 2266\;MHz, we report an execution time between 500\;ns and 150\;ns, which confirms its negligible impact on system performance.

\begin{table}[t]
    \centering
    \caption{Energy estimation error of the system-level power model at different frequencies over the full validation set}
    \label{tab:comb_energy}
    \begin{tabular}{@{}lZ{2pt}lZ{7pt}cccZ{7pt}cccZ{7pt}ccc@{}}
        \toprule
        \multirow{2}{*}{\textbf{\begin{tabular}[c]{@{}c@{}}Freq\\ {[}MHz{]}\end{tabular}}} & &  \footnotesize{\textbf{CPU}} &  & \multicolumn{3}{c}{730} &  & \multicolumn{3}{c}{1190} & &  \multicolumn{3}{c}{2266} \\
          & & \footnotesize{\textbf{GPU}} & &  \footnotesize{319} & \footnotesize{829} & \footnotesize{1377} & &  \footnotesize{319} & \footnotesize{829} & \footnotesize{1377} & &  \footnotesize{319} & \footnotesize{829} & \footnotesize{1377} \\ \midrule
         & & \footnotesize{\textbf{Error}} & &  \footnotesize{5.7\%} & \footnotesize{1.8\%} & \footnotesize{1.4\%} & &  \footnotesize{1.8\%} & \footnotesize{1.5\%} & \footnotesize{0.7\%} & &  \footnotesize{6.6\%} & \footnotesize{1\%} & \footnotesize{0.6\%} \\
        \bottomrule
    \end{tabular}
\end{table}


\section{Conclusions}\label{sec:concl}

In this work, we described a systematic, data-driven approach to \gls{dvfs}-aware statistical power modeling of the CPU and GPU sub-systems of a \gls{dvfs}-enabled heterogeneous platform using their respective \glspl{pmc}. 
Our proposed methodology achieves an unprecedented combination of general applicability,
automated model construction, lightweight model evaluation, and high accuracy.
We tackled the mentioned challenges with a data-driven statistical methodology for the model parameters selection, a \gls{lut}-based nature for the system-level model, and the linearity of the individual per-sub-system and per-frequency power models.
The validation of our power models on the NVIDIA Jetson AGX Xavier reported an energy estimation accuracy aligned or superior to the \gls{soa} while achieving desirable responsiveness and decomposability, complemented by a low model evaluation runtime.
These results pave the way for further work to assess the benefits of our models to online power monitoring and \gls{dpm} policies based on online power estimates. Further investigations will also include the application of our approach to different target platforms for the quantification of its general applicability.


\section*{Acknowledgements}
Supported in part by the European Union's Horizon 2020 research and innovation program, in the context of the AMPERE (\#871669) and Fractal (\#877056) projects.


%
\bibliographystyle{splncs04}
\bibliography{paper}

\begin{thebibliography}{10}
\providecommand{\url}[1]{\texttt{#1}}
\providecommand{\urlprefix}{URL }
\providecommand{\doi}[1]{https://doi.org/#1}

\bibitem{ahmad2017survey}
Ahmad, R.W., et~al.: A survey on energy estimation and power modeling schemes
  for smartphone applications. International Journal of Communication Systems
  \textbf{30}(11),  e3234 (2017)

\bibitem{amdgpuperfapi}
{AMD GPUOpen}: {GPUPerfAPI} v3.10 user guide (2021),
  \url{https://gpuperfapi.readthedocs.io/en/latest/index.html}

\bibitem{armcortexA57}
{ARM Holdings}: {ARM Cortex-A57 MPCore} processor technical reference manual
  (feb 2016), \url{https://developer.arm.com/documentation/ddi0488}

\bibitem{asanovic2006landscape}
Asanovic, K., et~al.: The landscape of parallel computing research: A view from
  berkeley  (2006)

\bibitem{bellosa2000benefits}
Bellosa, F.: The benefits of event: driven energy accounting in power-sensitive
  systems. In: Proceedings of the 9th workshop on ACM SIGOPS European workshop:
  beyond the PC: new challenges for the operating system. pp. 37--42 (2000)

\bibitem{bertran2012systematic}
Bertran, R., Gonzalez, M., Martorell, X., Navarro, N., Ayguade, E.: A
  systematic methodology to generate decomposable and responsive power models
  for {CMPs}. IEEE Transactions on Computers  \textbf{62}(7),  1289--1302
  (2012)

\bibitem{bertran2013counter}
Bertran, R., Gonzalez, M., Martorell, X., Navarro, N., Ayguad{\'e}, E.:
  Counter-based power modeling methods: Top-down vs. bottom-up. The Computer
  Journal  \textbf{56}(2),  198--213 (2013)

\bibitem{bircher2011complete}
Bircher, W.L., John, L.K.: Complete system power estimation using processor
  performance events. IEEE Transactions on Computers  \textbf{61}(4),  563--577
  (2011)

\bibitem{che2009rodinia}
Che, S., et~al.: Rodinia: A benchmark suite for heterogeneous computing. In:
  2009 IEEE international symposium on workload characterization (IISWC). pp.
  44--54. IEEE (2009)

\bibitem{isci2003runtime}
Isci, C., Martonosi, M.: Runtime power monitoring in high-end processors:
  Methodology and empirical data. In: Proceedings. 36th Annual IEEE/ACM
  International Symposium on Microarchitecture, 2003. MICRO-36. pp. 93--104.
  IEEE (2003)

\bibitem{lin2020taxonomy}
Lin, W., et~al.: A taxonomy and survey of power models and power modeling for
  cloud servers. ACM Computing Surveys (CSUR)  \textbf{53}(5),  1--41 (2020)

\bibitem{malony2011parallel}
Malony, A.D., et~al.: Parallel performance measurement of heterogeneous
  parallel systems with {GPUs}. In: 2011 international conference on parallel
  processing. pp. 176--185. IEEE (2011)

\bibitem{mammeri2019performance}
Mammeri, N., Neu, M., Lal, S., Juurlink, B.: Performance counters based power
  modeling of mobile {GPUs} using deep learning. In: 2019 International
  Conference on High Performance Computing \& Simulation (HPCS). pp. 193--200.
  IEEE (2019)

\bibitem{may2001mpx}
May, J.M.: {MPX}: Software for multiplexing hardware performance counters in
  multithreaded programs. In: Proceedings 15th International Parallel and
  Distributed Processing Symposium. IPDPS 2001. pp. 8--pp. IEEE (2001)

\bibitem{nvidiaxavier}
{NVIDIA Corporation}: {Jetson AGX Xavier} developer kit (2018),
  \url{https://developer.nvidia.com/embedded/jetson-agx-xavier-developer-kit}

\bibitem{nvidiacupti}
{NVIDIA Corporation}: {CUPTI} v11.6 user guide (2021),
  \url{https://docs.nvidia.com/cupti}

\bibitem{pi2019study}
Pi~Puig, M., De~Giusti, L.C., Naiouf, M., De~Giusti, A.E.: A study of hardware
  performance counters selection for cross architectural {GPU} power modeling.
  In: XXV Congreso Argentino de Ciencias de la Computaci{\'o}n
  (CACIC)(Universidad Nacional de R{\'\i}o Cuarto, C{\'o}rdoba, 14 al 18 de
  octubre de 2019) (2019)

\bibitem{pusukuri2009methodology}
Pusukuri, K.K., Vengerov, D., Fedorova, A.: A methodology for developing simple
  and robust power models using performance monitoring events. proceedings of
  WIOSCA  \textbf{9} (2009)

\bibitem{singh2009real}
Singh, K., Bhadauria, M., McKee, S.A.: Real time power estimation and thread
  scheduling via performance counters. ACM SIGARCH Computer Architecture News
  \textbf{37}(2),  46--55 (2009)

\bibitem{taylor2012dark}
Taylor, M.B.: Is dark silicon useful? harnessing the four horsemen of the
  coming dark silicon apocalypse. In: DAC Design Automation Conference 2012.
  pp. 1131--1136. IEEE (2012)

\bibitem{walker2016accurate}
Walker, M.J., et~al.: Accurate and stable run-time power modeling for mobile
  and embedded {CPUs}. IEEE Transactions on Computer-Aided Design of Integrated
  Circuits and Systems  \textbf{36}(1),  106--119 (2016)

\bibitem{wang2019statistic}
Wang, Q., Li, N., Shen, L., Wang, Z.: A statistic approach for power analysis
  of integrated {GPU}. Soft Computing  \textbf{23}(3),  827--836 (2019)

\end{thebibliography}

\end{document}